\documentclass{osa-article}

\journal{osajournal}

\articletype{Research Article}

\begin{document}

\title{Modelling and optimization of the excitonic diffraction grating}

\author{P.Yu. Shapochkin, Yu.V. Petrov, S.A. Eliseev, V.A. Lovcjus,  Yu.P. Efimov, and Yu.V. Kapitonov\authormark{*}}

\address{Saint-Petersburg State University, ul. Ulyanovskaya 1, St. Petersburg 198504, Russia}

\email{\authormark{*}yury.kapitonov@spbu.ru} 


\begin{abstract}
Periodical spatial modulation of the excitonic resonance in a quantum well could lead to the formation of a new highly directional and resonant coherent optical response -- resonant diffraction. Such excitonic diffraction gratings were demonstrated in epitaxially grown quantum wells patterned by the low-dose ion beam irradiation before or after the growth. In this paper we present a theoretical model of the resonant diffraction formation based on the step-by-step approximation of the Maxwell equation solution. The resulting theory allows us to reliably describe experimental data, as well as to predict the way to increase the diffraction efficiency.
\end{abstract}

\section{Introduction}
Semiconductor quantum wells (QWs) possessing excitonic resonance are key elements that underlie many modern and proposed information photonics devices, including optical memory~\cite{Poltavtsev2014}, polariton lasers~\cite{Savvidis2000} and quantum simulators~\cite{Amo2016} based on the exciton-polaritons in Bragg microcavities~\cite{Kavokin2008}, polaritonic circuits utilizing propagating exciton-polaritons in planar waveguides~\cite{Walker2013,Shapochkin2018} and many others. In these devices quantum wells are either kept uniform in lateral direction, or the sample is patterned as a whole, i.e. by etching of micropillars~\cite{Kalevich2015}.

There are methods that could lead to the spatial modulation of excitonic properties of QWs without change of non-resonant properties of the sample (surface relief, roughness, or background refractive index). One of such methods is the ion beam induced intermixing~\cite{Allard1992}. Irradiation of a QW by an focused ion beam with ion dose below the milling threshold (usually $< 10^{16}$~ions/cm$^2$) could lead to the intermixing of QW heterointerfaces. Smoothing of the QW confinement profile results in the blue shift of the exciton resonance. Irradiated QW optical quality could be restored to some extent by the rapid thermal annealing~\cite{Poole1994}. The defects formation could be used as a spatial modulation itself. At much lower doses ($<10^{13}$~ions/cm$^2$) no intermixing effects take place. Despite the fact that this defect formation leads to the photoluminescence quenching by opening of new non-radiative recombination pathways~\cite{Linfield1990}, its main effect on the coherent optical response of QW is an inhomogeneous broadening of the excitonic resonance~\cite{Kapitonov2015}. For the most common Ga$^+$ ion beam incident on the GaAs crystal the ion-solid interaction volume will have radius around 50~nm. Thus the sub-wavelength QW modulation is possible, and QW-based diffractive optical elements could be made.

In~\cite{Kapitonov2016} we have utilized the post-molecular-beam-epitaxy (MBE) growth ion beam irradiation for spatial modulation of the inhomogeneous broadening of the exciton resonance in QW. Periodical spatial modulation lead to the formation of a new coherent response -- resonant diffraction. Laser beam diffracted by this diffraction grating has high directionality, but the diffraction efficiency spectrum demonstrates very sharp resonance without any background since only the light with the energy close to the excitonic resonance is diffracted. By analogy with the reflection of light from a uniform quantum well (''excitonic mirror''), for such an object we will have proposed the term ''excitonic diffraction grating''.

For a theoretical description of this phenomenon in~\cite{Kapitonov2016} we used the single scattering model. This model allowed us to obtain a qualitative description of the reflection and diffraction spectra. However, this model is applicable only in the case of the radiative width of the excitonic resonance $\Gamma_R$ being much smaller than the non-radiative broadening $\Gamma_{NR}$. For QW used in~~\cite{Kapitonov2016} $\frac{\Gamma_R}{\Gamma_{NR}} \approx 0.3$. For higher quality molecular beam epitaxy grown QWs this ratio could reach 0.45 for GaAs/Al$_{0.3}$Ga$_{0.7}$As QWs~\cite{Poltavtsev2008}, 0.6 for GaAs/Al$_{0.03}$Ga$_{0.97}$As QWs~\cite{Solovev2017}. It also could exceed unity for thin~\cite{Poltavtsev2014_2} and thick~\cite{Trifonov2015} In$_{x}$Ga$_{1-x}$As/GaAs ($x<0.1$) QWs.

In this work we present a theoretical modeling of the excitonic diffraction grating by the step-by-step approximation of the Maxwell equation solution. This model, being applicable for large $\frac{\Gamma_R}{\Gamma_{NR}}$, makes it possible to correctly describe experimental data presented in~~\cite{Kapitonov2016}, as well as in our earlier work~\cite{Kapitonov2013}, where resonant diffraction was observed from the QW grown on the substrate, pre-patterned by the ion beam irradiation. Developed model was also used to show the way to the further increase of the diffraction efficiency, and to predict resonant diffraction spectra in the case of the spatial modulation of the excitonic resonance radiative width or spectral position.

\section{Theoretical consideration}

\subsection{Formulation of the problem}

Let us consider the problem of the light scattering on a heterostructure with a spatially modulated QW consisting from three regions perpendicular to the growth axis $z$: half-space I with the refractive index $n_1$, and layer II with thickness $h$ and half-space III with refractive indexes $n_2$ separated by the infinitely thin QW with the susceptibility $\tilde g(x,\omega)$, where $x$ -- coordinate in the direction of modulation, $\omega$ -- light energy. Suppose the structure is infinite in directions $x$ and $y$.

From the upper half-space, a linearly polarized plane wave with light energy~$\omega$ falls on the layer~II with the angle of incidence~$\theta_1$. The module of the wave vector of light is $k = \frac{\omega}{\hbar c}$. We will find the solution for the case of the light polarized in the plane of incidence (TM-polarization). Similar analysis could be made for the TE-polarized light. We assume excitation in the linear mode, and $k$ and phase advance in the layer~II independent  from~$\omega$. The $x$-projection of the incident light wave vector is $k_x = k n_j \sin \theta_j$, $j = 1, \, 2$. Conservation of $k_x$ on the I/II interface leads to the Snell's law $n_2 \sin \theta_2 = n_1 \sin \theta_1$, where $\theta_2$ -- angle of refraction. For TM-polarization there is a special case of the incidence at the Brewster angle $\theta_{Br} = \arctan \left( \frac{n_2}{n_1} \right)$, when the reflection from the I/II interface disappears.

\subsection{Maxwell equations solution}

    We wish to find the light field scattered in the upper half-space~I. We will represent it in the form of an expansion in $x$-components of the wavevector, which we will denote by~$q$, with coefficients $R(q)$. The $z$-projection of the expansion wavevector is $k_{zj}(q) =  \sqrt{k^2 n_j^2 - q^2}$, $j=1,2$.
    We will also consider descending and ascending fields in the layer~II and only descending (transmitted) field in the half-space~III. 
    We introduce the effective susceptibility $g(x) =  \frac{2 \pi k \cos \theta_2}{n_2} \tilde g(x)$. We represent $g(x)$ as a Fourier expansion $g(x) = \int G(q) e^{i q x} \, d q$. The phase advance experienced by the light passing in one direction through the layer~II will be $\varphi(q) = hk_{z2}(q)$. The phase advance of the refracted light we will denote as $\varphi = \varphi(k_x) = h k n_2 \cos \theta_2$. We introduce the following expression:

	\begin{equation}	
	r_{\pm}(q,q') =
	\frac{
	   n_2^2 k_{z1}(q') \pm n_1^2 k_{z2}(q')
	   }
	   {
	    n_2^2 k_{z1}(q) + n_1^2 k_{z2}(q) 
	   }
	   \cdot
	\frac{
	    k_{z2}(q)
	    }
	    {
	    k_{z2}(q')
	    }.
	\end{equation}		
	
	We will also denote a simple Fresnel expression for the reflection coefficient as $r = r_{-}(k_x,k_x)$. Light incident at the Brewster angle $\theta_1 = \theta_{Br}$ corresponds to $r = 0$. We also introduce the interference factor:
	
	\begin{equation}	
	a_{\pm}(q,q') =
	e^{i( \varphi(q) + \varphi(q') )}
	\left(
		r_{\mp}(q,q') + r_{\pm}(q,q') e^{-2i \varphi(q')}
	\right).
	\end{equation}	
	

    Accounting for boundary conditions for the I/II and II/III interfaces leads to the following integral equation, allowing to find the distribution of the scattered field $R(q)$ by known susceptibility Fourier expansion $G(q)$:

	\begin{equation}
	\label{RqEq}	
	R(q) -
	i \int 
	\underbrace{a_-(q,q') \frac{k_{z2}(q')}{k_{z2}(k_x)} G(q-q') }_{{\bf K}(q,q')} R(q')
	\, d q'		
	=
	r \delta (q-k_x) -	i a_+(q,k_x) G(q-k_x).
	\end{equation}
	
	This is the Fredholm integral equation of the second kind with the kernel ${\bf K}(q,q')$, which cannot be represented as ${\bf K}(q-q')$ or ${\bf K_1}(q) {\bf K_2}(q')$. We note that at this stage no assumptions were made about $G(q)$.

\subsection{Expansion of the exact solution}

	Consider effective susceptibility $g(x)$ modulated periodically in the $x$-direction with period $L$. We assume that the amplitude of the spatial modulation is small by introducing a small parameter $\epsilon$ in front of all expansion terms except zeroth:
	
	\begin{equation}
    \label{gexpand}
	G(q)
	=
	G_0 \delta(q) + \epsilon \sum^{+\infty}_{n=-\infty, n \ne 0} G_n \delta \left(q+q_n \right),
	\end{equation}	
	
	where $q_n=\frac{2 \pi n}{L}$. We search scattered field $R(q)$ as an expansion in small parameter $\epsilon$: $R(q) = \sum_{m=0}^{+\infty} \epsilon^m R_{(m)}(q)$, where the index in parentheses denotes the expansion term number. After substitution of the expansions for $G(q)$ and $R(q)$ into the equation~(\ref{RqEq}) we get the following expression:

	\begin{equation}
	R(q) = \sum_{p=0}^{+\infty} 
	\sum^{+\infty}_{m=-\infty}
	\epsilon^p R_{(p)m} \delta(q - k_x + q_m),
	\end{equation}
	
	where $\delta(x)$ -- Dirac delta function, and:
	
	\begin{equation}	
	\label{R0}
	R_{(0)0}
	=
	\frac{r - i r G_0 - i G_0 e^{2i\varphi}}
	{1 - i G_0 - i r G_0 e^{2 i\varphi}},
	\qquad	
	R_{(1)n}
	=
	\frac{
	i G_n
	\left(
	a_-(k_x-q_n,k_x) R_{(0)0}
	-
	a_+(k_x-q_n,k_x)
	\right)
	}
	{1 - i a_-(k_x-q_n,k_x-q_n) \frac{k_{z2}(k_x-q_n)}{k_{z2}(k_x)} G_0},
	\end{equation}	

    and coefficients for $p>1$ could be calculated using the following recurrent expression for $p$-terms of the $n$-diffraction reflex:
		
	\begin{equation}	
	R_{(p)n}
	= 
	\sum^{+\infty}_{j=-\infty, j \ne 0}
	\frac{i a_-(k_x - q_{n},k_x - q_{n-j}) \frac{k_{z2}(k_x - q_{n-j})}{k_{z2}(k_x)}  G_j}
	{1 - i a_-(k_x - q_{n},k_x - q_{n}) \frac{k_{z2}(k_x - q_{n})}{k_{z2}(k_x)} G_0}
	R_{(p-1)n-j}.
	\end{equation}	

	This expansion can be used for a step-by-step approximation of spectral dependencies of the reflection coefficient and diffraction efficiencies. One could see that in this expansion there are no directions other than those determined by the diffraction grating equation $q = k_x - q_n$. Each subsequent expansion term is calculated based on the values of \textit{all} coefficients of the previous term. The term $R_{(0)0}$ corresponds to the reflection from an excitonic mirror with $g(x)= G_0$, and corrections for reflection occur only for the terms of the expansion $\sim \epsilon^2$ and further. For high order diffraction reflexes one should consider the possibility of the total internal reflection at I/II interface.

    Next we will consider the case, when the light is incident at the Brewster angle. We will also assume small diffraction angles ($q_n \rightarrow 0$). In this case coefficients (\ref{R0}) take simple form:

	\begin{equation}	
	R_{(0)0}
	=
	\frac{-i G_0}{1 - i G_0} e^{2i\varphi}
	, \qquad
	R_{(1)n}
	=
	\frac{-i G_n}{(1 - i G_0)^2} e^{2i\varphi}
	.
	\end{equation}	

    Reflection coefficient and diffraction efficiencies could be calculated as $K_0 = \left| R_0 \right|^2$ and $K_n = \left| R_n \right|^2$ for the specified QW susceptibility $g(x)$.

\subsection{Quantum well susceptibility}
	
	Consider a QW with a periodic piecewise modulation of its properties of along the $x$-axis represented by the following susceptibility ($m$ -- integer number):

    \begin{equation}
	g(x) =
    \left\{
    \begin{array}{ll}
    	g_1, & x \in (mL, \, mL + l];\\
        g_2, & x \in (mL + l, \, (m+1) L].
    \end{array}
    \right.
    \end{equation}	
    
    For the susceptibility given in this way, the coefficients of the Fourier expansion~(\ref{gexpand}) are defined as follows (we introduce the fill factor $\alpha = \frac{l}{L}$):

	\begin{equation}
	G_n =
	\left\{
	\begin{array}{lcll}
		{\displaystyle \alpha g_1 + (1 - \alpha) g_2, } & n = 0,
	\\	
		\mathstrut
	\\
		{\displaystyle \frac{ \sin \left( \pi n \alpha \right)}{\pi n} e^{i \pi n \alpha} (g_1 - g_2), } & n \neq 0.
	\end{array}
	\right.
	\end{equation}

    The local susceptibility of a thin QW could be represented as $	g_{1,2} = \frac{\Gamma_{R1,2}}{\Delta \omega_{1,2} - i \Gamma_{NR1,2}},	$
    where $\Delta \omega_{1,2} = \omega_{1,2} - \omega$ is detuning of the excitonic resonance centered at energy $\omega_{1,2}$ from incident light energy, $\Gamma_{R1,2}$ --- radiative widths and $\Gamma_{NR1,2}$ --- nonradiative broadenings of the excitonic resonances of corresponding grating grooves. The nonradiative broadening consist of reversible (resulting from the inhomogeneous broadening) and irreversible phase relaxation rates.
    
    Next we will calculate $K_0(\Delta \omega)$ and $K_n(\Delta \omega)$ for different types of the QW modulation. For simplicity we will assume $\alpha = \frac{1}{2}$.

\subsection{Absence of the modulation}

    The simplest case is the absence of the modulation ($\Delta \omega_{1} = \Delta \omega_{2} = \Delta \omega$, $\Gamma_{R1} = \Gamma_{R2} = \Gamma_{R}$, $\Gamma_{NR1} = \Gamma_{NR2} = \Gamma_{NR}$). In this case, there is no diffraction ($K_n(\Delta \omega) = 0$), and the reflection spectrum is described by a known result for an ''excitonic mirror''~\cite{Poltavtsev2008, Poltavtsev2014_2}:

    \begin{equation}
    \label{K0}
	K_0(\Delta \omega) = \frac{\Gamma_R^2}{\Delta \omega^2  + \left( \Gamma_R + \Gamma_{NR} \right)^2}.
	\end{equation}    
	
	Reflection spectrum has a form of an Lorentian curve centred at $\Delta \omega = 0$ with resonant reflection coefficient $K_0(0) = \frac{\Gamma_R^2}{\left( \Gamma_R + \Gamma_{NR} \right)^2}$, and half width at half maximum (HWHM) equal to $\Gamma_R + \Gamma_{NR}$.

\subsection{Modulation of the radiative width}

    Let us consider the modulation of the radiative width of the excitonic resonance ($\Delta \omega_{1} = \Delta \omega_{2} = \Delta \omega$, $\Gamma_{NR1} = \Gamma_{NR2} = \Gamma_{NR}$). We will introduce average radiative width $\tilde{\Gamma}_R = \frac{\Gamma_{R1} + \Gamma_{R2}}{2}$. Reflection coefficient $K_0(\Delta \omega)$ is described by~(\ref{K0}) with the substitution $\Gamma_R \rightarrow \tilde{\Gamma}_R$. Diffraction efficiency will be the following:
 
 	\begin{equation}
	K_1(\Delta \omega) =
    	\frac{1}{\pi^2}
			\cdot
			\frac{
			\left(
			    \Gamma_{R1}-\Gamma_{R2}
			\right)^2
			\left(
			    \Delta \omega^2+\Gamma_{NR}^2
			\right)
			}{
			\left(
			    \Delta \omega^2+
			        \left(
		                \tilde{\Gamma}_R + \Gamma_{NR}
			        \right)^2
			\right)^2
			}.
	\end{equation}   
    
    Far from the resonance, this function behaves like a Lorentzian $\sim \frac{1}{
			    \Delta \omega^2+
			        \left(
		                \tilde{\Gamma}_R + \Gamma_{NR}
			        \right)^2}$.
	More curious is the diffraction behavior near the resonance. For low-quality quantum wells, the spectrum has the form of a peak. However, in the case of $\frac{\tilde{\Gamma}_R}{\Gamma_{NR}} > \sqrt{2} - 1$, the diffraction spectrum splits into two symmetric components with the splitting magnitude equal to $2 \sqrt{ \tilde{\Gamma}_R^2 + 2\tilde{\Gamma}_R \Gamma_{NR} - \Gamma_{NR}^2 }$. In the limit of $\Gamma_{NR} \rightarrow 0$ splitting magnitude reaches $2 \tilde{\Gamma}_R$ and diffraction efficiency at the resonance falls to zero. This can be explained by the fact that in the absence of nonradiative broadening, the reflection from even and odd grooves of the diffraction grating reaches unity at the resonance, which leads to the disappearance of the diffraction.
	
	Figure 2~(a,b) shows the theoretical reflection and diffraction spectra for a grating with grooves $\Gamma_{R1} = 40$~$\mu$eV and $\Gamma_{R2} = 40$~$\mu$eV and $\alpha = \frac{1}{2}$ for various $\Gamma_{NR}$. At $\Gamma_{NR} < \frac{\tilde{\Gamma}_R}{\sqrt{2} - 1} \approx 72.4$~$\mu$eV the diffraction spectrum  splitting is observed.

\subsection{Modulation of the resonance frequency}

    Let us consider the modulation of the excitonic resonance frequency ($\Gamma_{R1} = \Gamma_{R2} = \Gamma_{R}$, $\Gamma_{NR1} = \Gamma_{NR2} = \Gamma_{NR}$). Let us make the following change of variables: $\Delta \omega_{1,2} = \Delta \omega \pm \delta$. In this case the result will be as follows:

 	\begin{equation}
	K_0(\Delta \omega) =
        \frac{
		    \Gamma_{R}^2
		}{
            \Delta \omega^2 +
            \left( \Gamma_R + \Gamma_{NR} \right)^2 -
            \frac{\delta^2 \left( \delta^2 + 2 \left( \Delta \omega^2 - \Gamma_{NR} \left( \Gamma_R + \Gamma_{NR} \right) \right) \right)}{ \Delta \omega^2 + \Gamma_{NR}^2}
		},
	\end{equation}   

 	\begin{equation}
	K_1(\Delta \omega) =
	\frac{1}{\pi^2}
	\cdot
        \frac{
		   4 \delta^2 \Gamma_{R}^2
		   \left( 1 + \frac{\delta (\delta - 2 \Delta \omega) }{ \Delta \omega^2 + \Gamma_{NR}^2 } \right)
		   \left( 1 + \frac{\delta (\delta + 2 \Delta \omega) }{ \Delta \omega^2 + \Gamma_{NR}^2 } \right)
		}{
            \left(
                \Delta \omega^2 +
                \left( \Gamma_R + \Gamma_{NR} \right)^2 -
                \frac{\delta^2 \left( \delta^2 + 2 \left( \Delta \omega^2 - \Gamma_{NR} \left( \Gamma_R + \Gamma_{NR} \right) \right) \right)}{ \Delta \omega^2 + \Gamma_{NR}^2}
            \right)^2
		},
	\end{equation}  

    In case of sufficiently large $\delta$, first the reflection spectrum, and then the diffraction spectrum, are split into two peaks. This is illustrated in Fig.2~(c,d). It shows the theoretical reflection and diffraction spectra for a grating with grooves having identical $\Gamma_R = 40$~$\mu$eV and $\Gamma_{NR} = 100$~$\mu$eV, but with different splittings~$\delta$ and $\alpha = \frac{1}{2}$. For even smaller $\Gamma_{NR}$, each of the resonances in the diffraction spectrum is split into two more.

\subsection{Modulation of the nonradiative broadening}

    Of greatest interest is the case of modulation of nonradiative broadening realized in experiments ($\Delta \omega_{1} = \Delta \omega_{2} = \Delta \omega$, $\Gamma_{R1} = \Gamma_{R2} = \Gamma_{R}$). In this case, the result will be as follows:
    
	\begin{equation}
	K_0(\Delta \omega) =
			\frac{
				\Gamma_R^2 \left(\Delta \omega^2 + \left( \frac{\Gamma_{NR1} + \Gamma_{NR2}}{2} \right)^2 \right)
			}{
				\left(\Delta \omega^2 + \tilde \Gamma_{1}^2 \right)
				\left(\Delta \omega^2 + \tilde \Gamma_{2}^2 \right)	
			},
    \end{equation}
	
	\begin{equation}
	\label{K1inh}
	K_1(\Delta \omega) =
    	\frac{1}{\pi^2}
			\cdot
			\frac{
				4 \Gamma_R^2 \left(\Gamma_{NR2} - \Gamma_{NR1} \right)^2
			    \left(\Delta \omega^2 + \Gamma_{NR1}^2 \right)
				\left(\Delta \omega^2 + \Gamma_{NR2}^2 \right)
			}{
				\left(\Delta \omega^2 + \tilde \Gamma_{1}^2 \right)^2
				\left(\Delta \omega^2 + \tilde \Gamma_{2}^2 \right)^2	
			},
	\end{equation}	

	where $\tilde \Gamma_{1,2}$ are introduced as follows:
	
	\begin{equation}
	\tilde{\Gamma}_{1,2} = 
	\frac{1}{\sqrt{2}}
	\sqrt{
		\Gamma_{NR1}^2 + \Gamma_{NR2}^2
		+
		(\Gamma_{NR1} + \Gamma_{NR2} + \Gamma_R)
		\left(\Gamma_R \pm \sqrt{(\Gamma_{NR2}-\Gamma_{NR1})^2 + \Gamma_R^2} \right)
	}.
	\end{equation}   

    Far from the resonance, the reflection coefficient behaves qualitatively as $K_0(\Delta \omega) \sim \frac{1}{\Delta \omega^2}$, similarly to~(\ref{K0}). The diffraction generated by the contrast of the spatial modulation decreases much faster: $K_1(\Delta \omega) \sim \frac{1}{\Delta \omega^4}$. For the same reason, the HWHM of the spectral peak $K_1(\Delta \omega)$ is less than the HWHM of $K_0(\Delta \omega)$ peak. Figure 2~(e,f) shows the theoretical reflection and diffraction spectra for a grating with grooves having identical $\Gamma_{R} = 40$~$\mu$eV. Nonradiative broadening of odd grating grooves in all cases is equal to $\Gamma_{NR1} = 100$~$\mu$eV, and $\Gamma_{NR2}$ is varied.

    As the sample temperature increases, the homogeneous broadening of the exciton resonance $\Gamma_2(T)$ associated with phonon scattering increases. Broadening leads to a decrease in the resonant reflection coefficient according to the law $K_0(0) \sim \frac{1} {\Gamma_2^2(T)} $. Simultaneously with the spectral broadening, the contrast of the diffraction grating is weakened, which leads to a more rapid decrease of the resonant diffraction efficiency according to the law $K_1(0) \sim \frac{1}{\Gamma_2^4(T)}$.
    
    \begin{figure}[ht]
    \centering\includegraphics[width=11 cm]{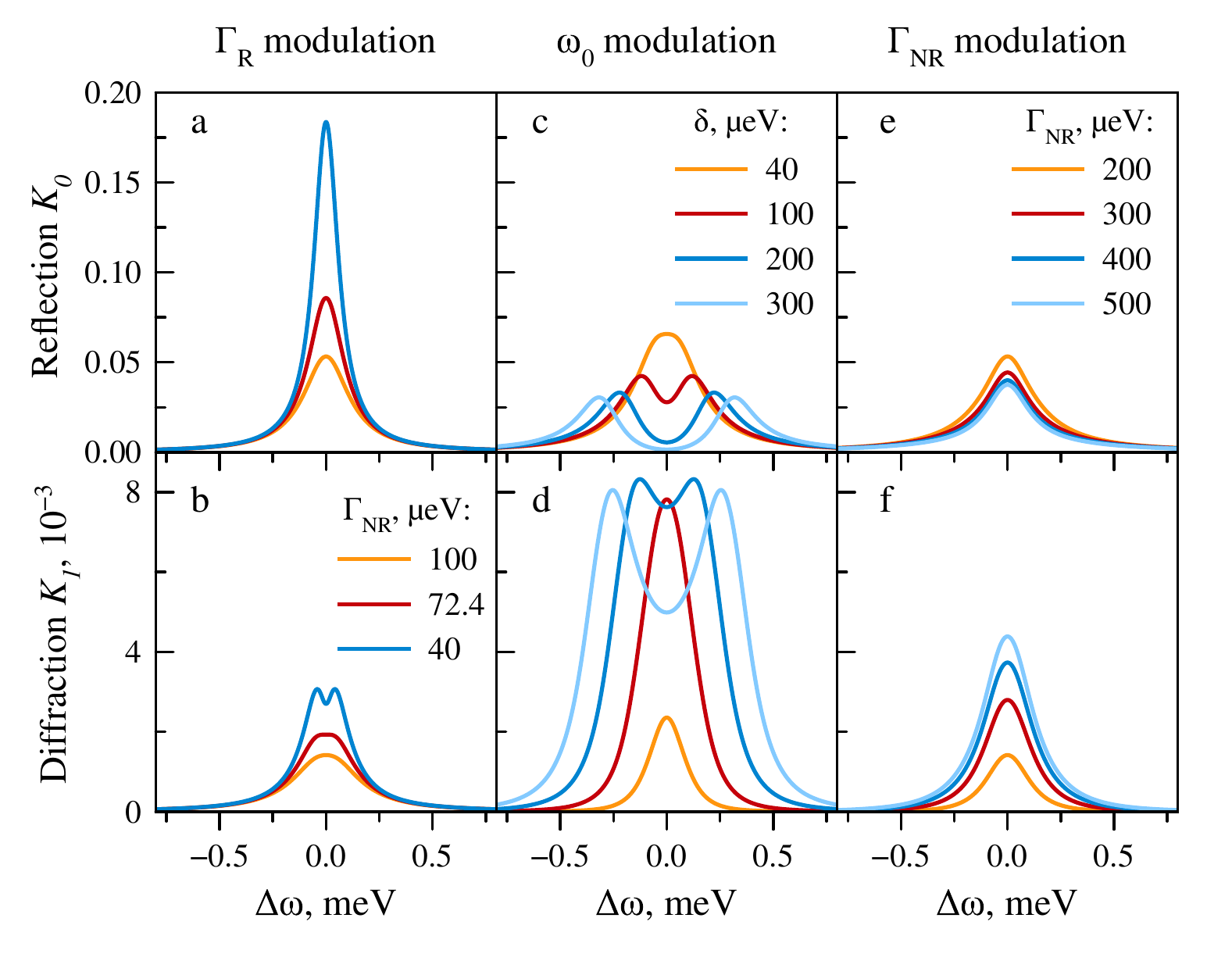}
    \caption{
    Theoretical spectral dependency of the reflection coefficient (a,c,e) and diffraction efficiency (b,d,f) for the case of the modulation of the radiative width $\Gamma_R$ (a,b) ($\Gamma_{R1} = 40$~$\mu$eV, $\Gamma_{R2} = 20$~$\mu$eV), resonance frequency $\omega_0$ (c,d) ($\Gamma_{R} = 40$~$\mu$eV, $\Gamma_{NR} = 100$~$\mu$eV), and nonradiative broadening $\Gamma_{NR}$ (e,f) ($\Gamma_{R} = 40$~$\mu$eV, $\Gamma_{NR1} = 100$~$\mu$eV). In all cases $\alpha = \frac{1}{2}$.}
    \label{Combined}
    \end{figure}

\subsection{Maximum diffraction efficiency}

    For practical applications of resonant diffraction gratings, an important issue is the conditions under which the maximum diffraction efficiency is achieved. Next, we consider the case of modulation of the nonradiative broadening. We find the resonant diffraction efficiency coefficient $K_1(0)$ in the same way as in~(\ref{K1inh}) but for an arbitrary $\alpha$:

	\begin{equation}
	K_1 (0) =
		\frac{\sin(\pi \alpha)^2}{\pi^2}
			\cdot
			\frac{
				\Gamma_R^2 \Gamma_{NR1}^2 \Gamma_{NR2}^2 (\Gamma_{NR2} - \Gamma_{NR1})^2
			}{
				(\Gamma_{NR1} \Gamma_{NR2} + \Gamma_R ( \Gamma_{NR1} (1 - \alpha) + \Gamma_{NR2} \alpha ) )^4	
			}.
	\end{equation}	

    The radiative width of the resonance $\Gamma_R$ and the non-radiative broadening of the unmodulated grating groove $\Gamma_{NR1}$ are determined by the QW growth conditions. Therefore these parameters can be considered fixed. Obviously, in this case, the maximum of the resonant diffraction efficiency is achieved at the maximum contrast between the grating grooves, i.e. at $\Gamma_ {NR2} \to +\infty$, and is equal to the following value (we denote $\gamma = \frac{\Gamma_{NR1}}{\Gamma_R}$):

	\begin{equation}
    	\label{gnrinfty}
	    \lim_{\Gamma_{NR2} \to + \infty} K_{1} (0) =
	    \frac{ \gamma^2 }{\pi^2}
	    \cdot
	    \frac{ \sin (\pi \alpha)^2 }{(\alpha + \gamma)^4 }.
	\end{equation}

    The value of $\alpha_{max} $, at which the maximum of this expression is reached, can be found as the zero of the first derivative with respect to $\alpha$, which leads to the following transcendental equation on $\alpha_{max}$:

	\begin{equation}
    	\label{transzeq}
	    \mathrm{cot} (\pi \alpha_{max}) = \frac{2}{\pi (\alpha_{max} + \gamma)}.
	\end{equation}
	
    Near $\alpha_{max} = \frac{1}{2}$, the cotangent can be approximated by the first two terms of the expansion series
    $
	    \mathrm{cot} (\pi \alpha_{max}) \approx \pi \left( \frac{1}{2} - \alpha_{max} \right).
	$
	Using this approximation, an approximate expression for $\alpha_{max}$ could be found:
	
	\begin{equation}
	    \alpha_{max} \approx
	    \frac{1 - 2 \gamma}{4}
	    +
	    \sqrt{\left( \frac{1 + 2 \gamma}{4} \right)^2 - \frac{2}{\pi^2}}.
	\end{equation}		 

\begin{figure}[ht]
    \centering\includegraphics[width=7cm]{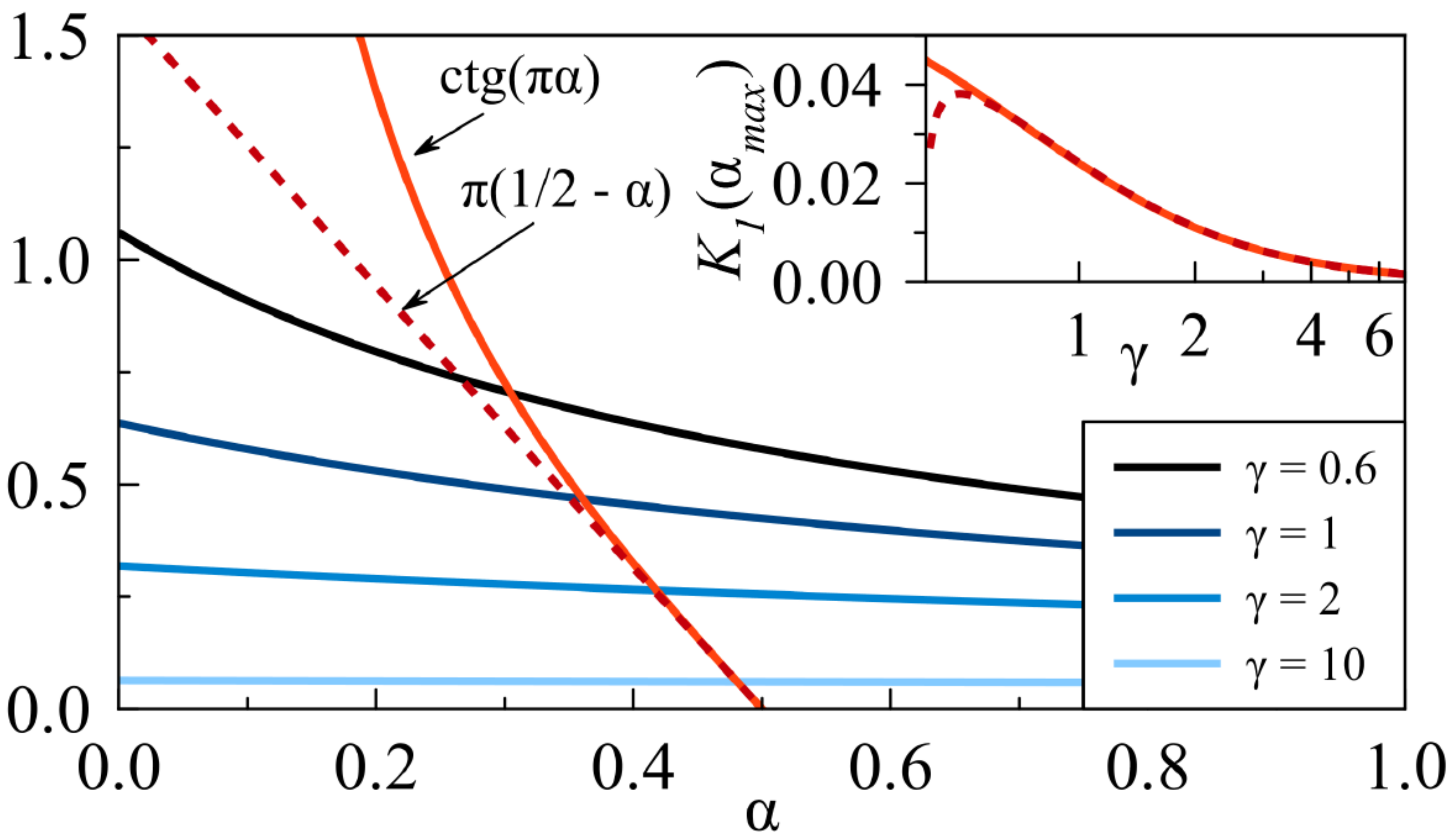}
    \caption{
    Graphic solution of the transcendental equation~(\ref{transzeq}). Cotangent (red solid line), its approximation (red dotted line) and the right side of the equation for different values of $\gamma$ (blue solid lines). Inset shows the dependence of the maximum resonant diffraction efficiency on $\gamma$ for $\alpha_{max}$, obtained by numerically solving the equation~(\ref{transzeq}) (solid line) and using approximation (dotted line).}
    \label{FigAlpha}
    \end{figure}

    Fig.~\ref{FigAlpha} shows the numerical solution of the equation~(\ref{transzeq}), and the proposed approximation. It can be seen that for sufficiently large $\gamma$, it is possible to use an approximating function to find the roots. For any $\gamma$, the maximum resonant diffraction efficiency is achieved when the modulated band is wider than the unmodulated ($\alpha_{max} \le \frac{1}{2}$). Figure inset shows the dependence of the maximum resonant diffraction efficiency on the parameter $\gamma$, obtained by substituting the root of the transcendental equation $\alpha_{max}$ into the expression~(\ref{gnrinfty}). It can be seen from the figure that for realistic values of $\gamma$ for GaAs-based QWs the maximum resonant diffraction efficiency in the Brewster geometry reaches 1~-~3\%.
	
\begin{figure}[ht]
    \centering\includegraphics[width=7cm]{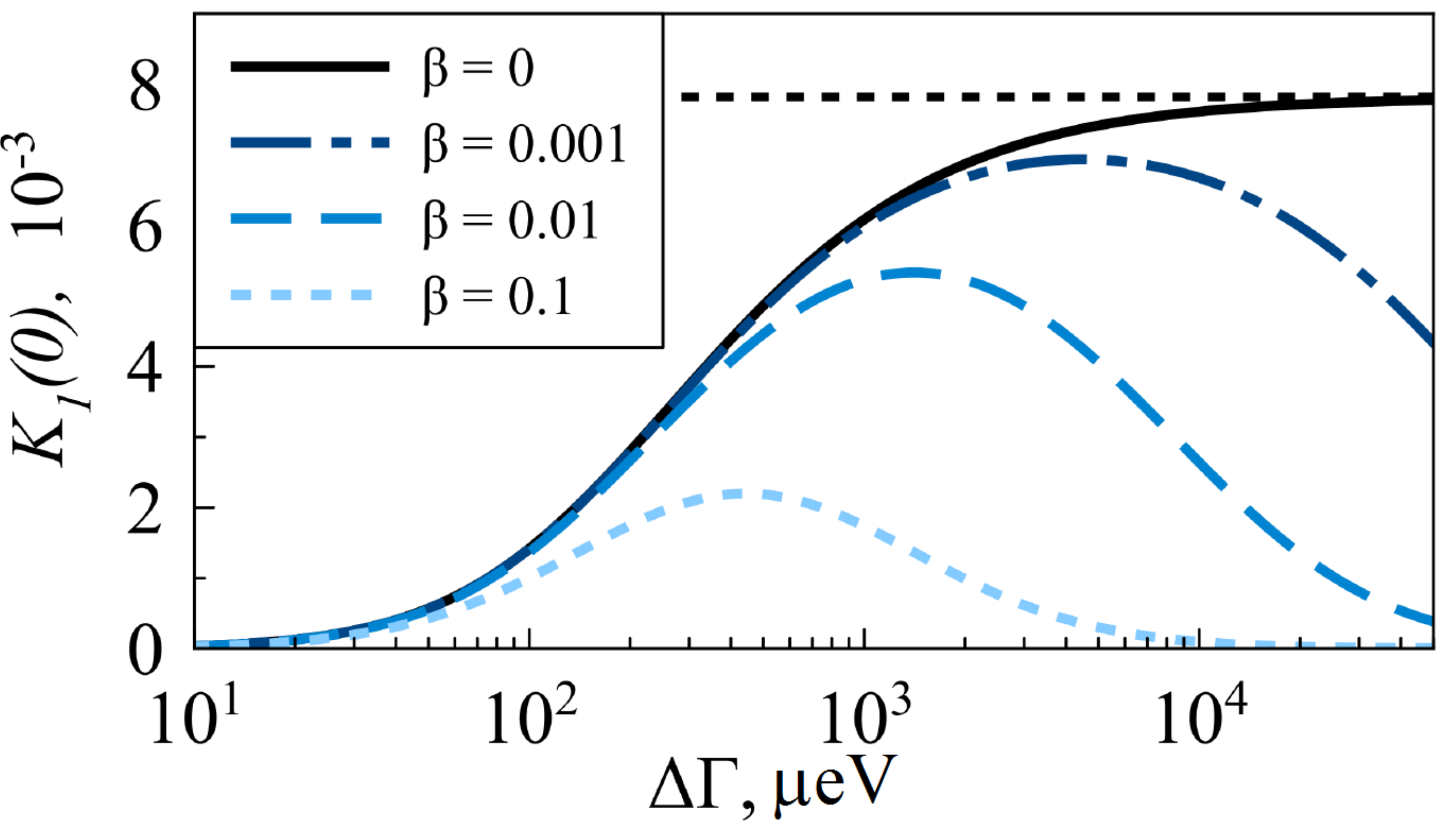}
    \caption{
    Resonant diffraction efficiency dependency on the modulation $\Delta \Gamma$ for different values of parasitic modulation coefficient $\beta$. The black dotted line shows the maximum achievable diffraction efficiency in accordance with~(\ref{gnrinfty}). The model parameters are $\Gamma_R = 40$~$\mu$eV, $\Gamma_{NR0} = 100$~$\mu$eV, $\alpha = \frac{1}{2}$.
    }
    \label{FigBeta}
    \end{figure}

    Above it was assumed that the nonradiative width of the unmodulated grating grooves $\Gamma_{NR1}$ does not depend on $\Gamma_{NR2}$. This assumption allowed us to make a transition $\Gamma_{NR2} \to +\infty$. This assumption is valid for sufficiently large $L$. However, for real modulation techniques a proximity effect takes place when the local modulation of a QW also leads to the modulation of adjacent QW regions. Such a modulation in the general case corresponds to a continuous function $g(x)$, obtained by convolving the ''instrumental response function'' with the modulation profile. This effect can be described qualitatively in the approximation of a piecewise constant $g(x)$ by the following substitution: $\Gamma_{NR1} = \Gamma_{NR0} + \beta \Delta \Gamma$, $\Gamma_{NR2} = \Gamma_{NR0} + \Delta \Gamma$, where $\Delta \Gamma $ is the modulation of the even grooves, and $\beta$ is the coefficient describing the parasitic modulation of odd grooves ($\beta < 1$). Fig.~\ref{FigBeta} shows the behavior of the resonant diffraction efficiency.

    In the absence of parasitic modulation ($\beta = 0$), the resonance diffraction efficiency asymptotically approaches the saturated value (dotted line in Fig.~\ref{FigBeta}). In the presence of parasitic modulation ($\beta > 0$), the dependence reaches a maximum at certain $\Delta \Gamma$. With a further increase of $\Delta \Gamma$, the overall  deterioration of the QW quality due to parasitic modulation leads to the decrease in the resonant diffraction efficiency to zero. As $\beta$ grows, the maximum achievable diffraction efficiency becomes smaller, and it is achieved at smaller $\Delta \Gamma$.

    Thus, for a QW with known $\Gamma_R$ and $\Gamma_{NR0}$, and known parameter of the parasitic modulation $\beta$, the above expressions allow one to find optimal modulation magnitude $\Delta \Gamma$ and the fill factor $\alpha$ at which the maximum resonant diffraction efficiency is achieved.

\section{Comparison with the experiment}

\subsection{Pre-MBE processing}

\begin{figure}[ht]
    \centering\includegraphics[width=7cm]{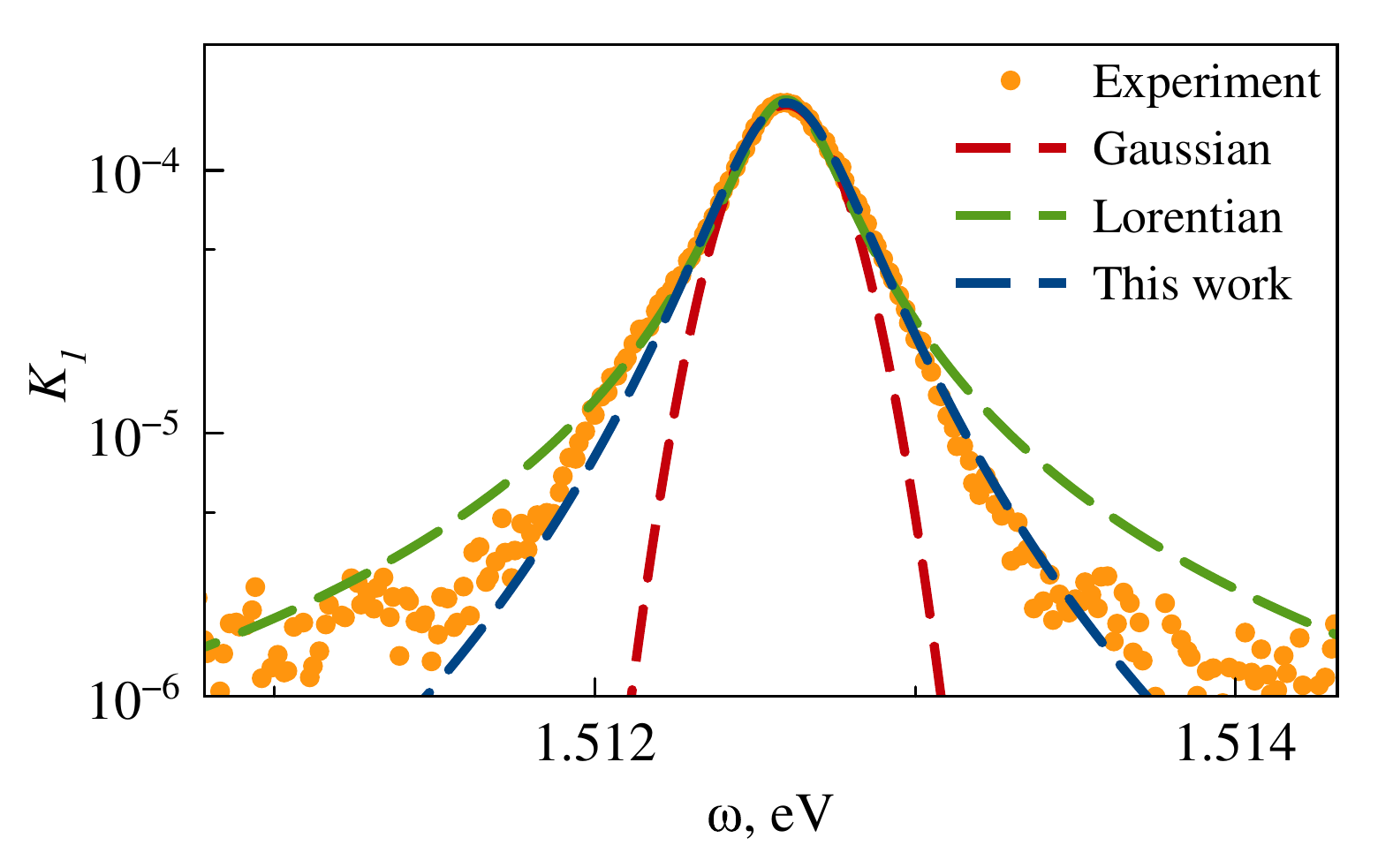}
    \caption{
    Experimental diffraction spectrum at 9.5~K for the pre-MBE processed sample (dots), its fitting by the Gaussian function (red dashed line), Lorentzian (green dashed line), and equation~(\ref{K1inh}) (blue dashed line). Note the logarithmic scale.
    }
    \label{FigP566_K1}
    \end{figure}

    Periodic spatial modulation of the inhomogeneous broadening of the exciton resonance can be obtained by growing QW on a substrate that was irradiated by an ion beam. In~\cite{Kapitonov2013} the In$_{0.02}$Ga$_{0.98}$As/GaAs quantum well separated by a 270~nm GaAs barrier from an irradiated GaAs substrate was studied. On the substrate, an array of lines with a period of 9~$\mu$m was irradiated with a 30~keV Ga$^+$ ion beam with dose 6.25$\cdot$~10$^9$ 1/cm$^2$. The experimental diffraction spectrum at 9.5~K for the first diffraction reflex $K_1(\omega)$ is shown in Fig.~\ref{FigP566_K1} by dots. More details on the optical experiment could be found in~~\cite{Kapitonov2013}. Far from the resonance, the spectrum decreases faster than Lorentzian $\sim \frac{1}{\Delta \omega^2 + 1}$, and more slowly than Gaussian function $\sim e^{-\Delta \omega^2}$. The spectrum is described most satisfactorily by the equation~(\ref{K1inh}).

\subsection{Post-MBE processing}
    
\begin{figure}[ht]
    \centering\includegraphics[width=11 cm]{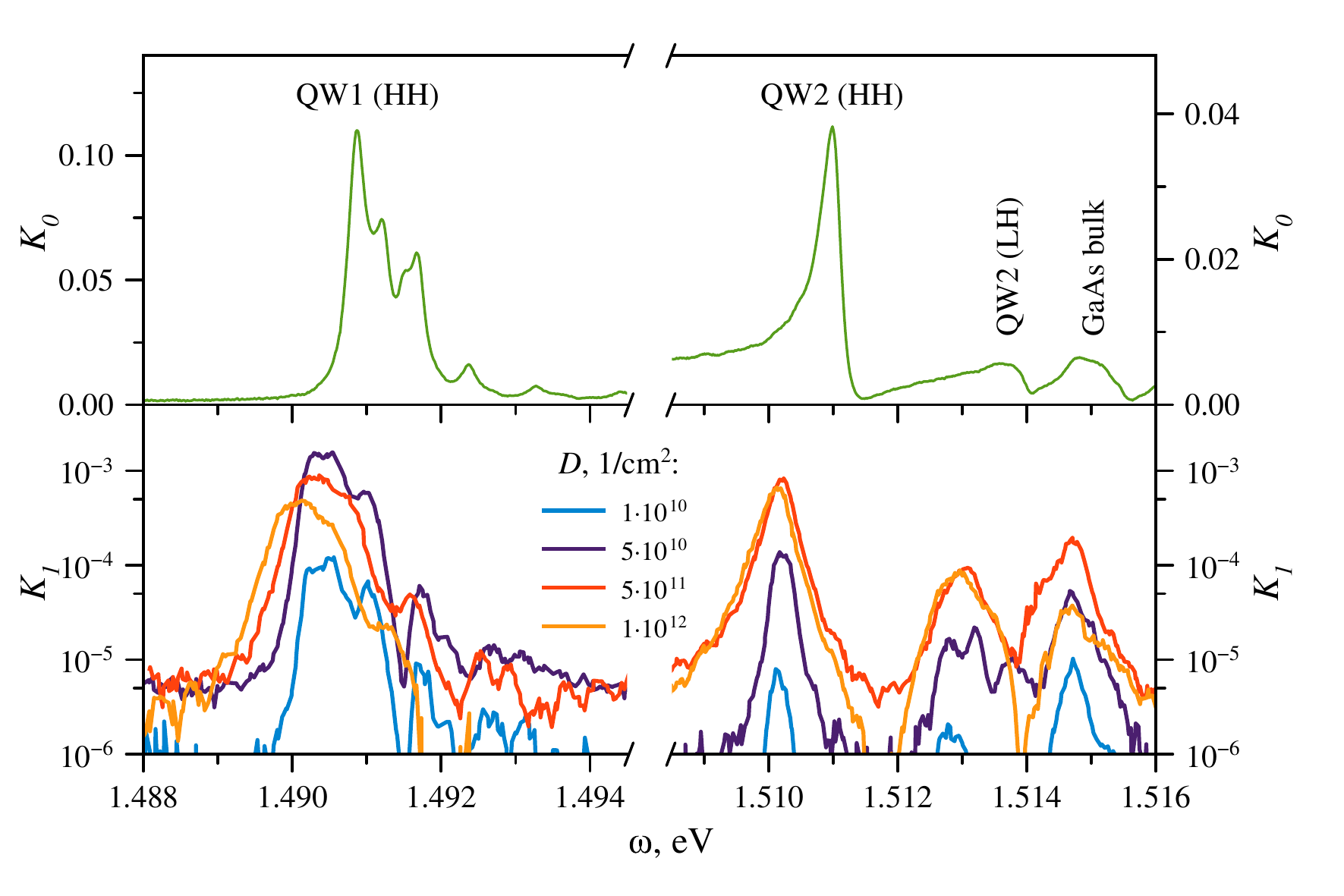}
    \caption{
    Reflection $K_0$ (top) and diffraction $K_1$ (bottom) spectra for a post-MBE processed QWs at 10~K. Diffraction spectra are shown for different ion irradiation doses $D$ in the logarithmic scale.
    }
    \label{FigP602_K0_K1}
    \end{figure}

    Another method of spatial modulation of exciton resonance inhomogeneous broadening is the irradiation of QWs by an ion beam after the MBE growth. In our previous work~\cite{Kapitonov2016} In$_{0.015}$Ga$_{0.985}$As/GaAs QWs were irradiated with a 35~keV He$^+$ ion beam. The sample has two QWs: thick QW1 with 190~nm width located 314.5~nm below the sample surface, and thin QW2 with 4.5~nm width 60~nm below the surface. Ion beam irradiation patterns represented periodic arrays of 400~nm width grooves with a period of 800~nm. Irradiation doses were from 5$\cdot$10$^{10}$ to 1$\cdot$10$^{12}$~1/cm$^2$. Fig.\ref{FigP602_K0_K1} shows the reflection (top) and diffraction (bottom) spectra at 10~K. A slight energy shift between spectra is due to the sample gradient. In spectra starting from 1.490~eV, features associated with the exciton quantization in a thick QW are observed ~\cite{Trifonov2015}. The ground state of a heavy-hole exciton is designated QW1~(HH).  At 1.510~eV and 1.513~eV are located the resonances of heavy- (HH) and light-hole (LH) excitons in the QW2 quantum well respectively. At 1.515~eV the resonance of the 3D-exciton in GaAs barrier is located (denoted as GaAs~bulk).
 
\begin{figure}[ht]
    \centering\includegraphics[width=6cm]{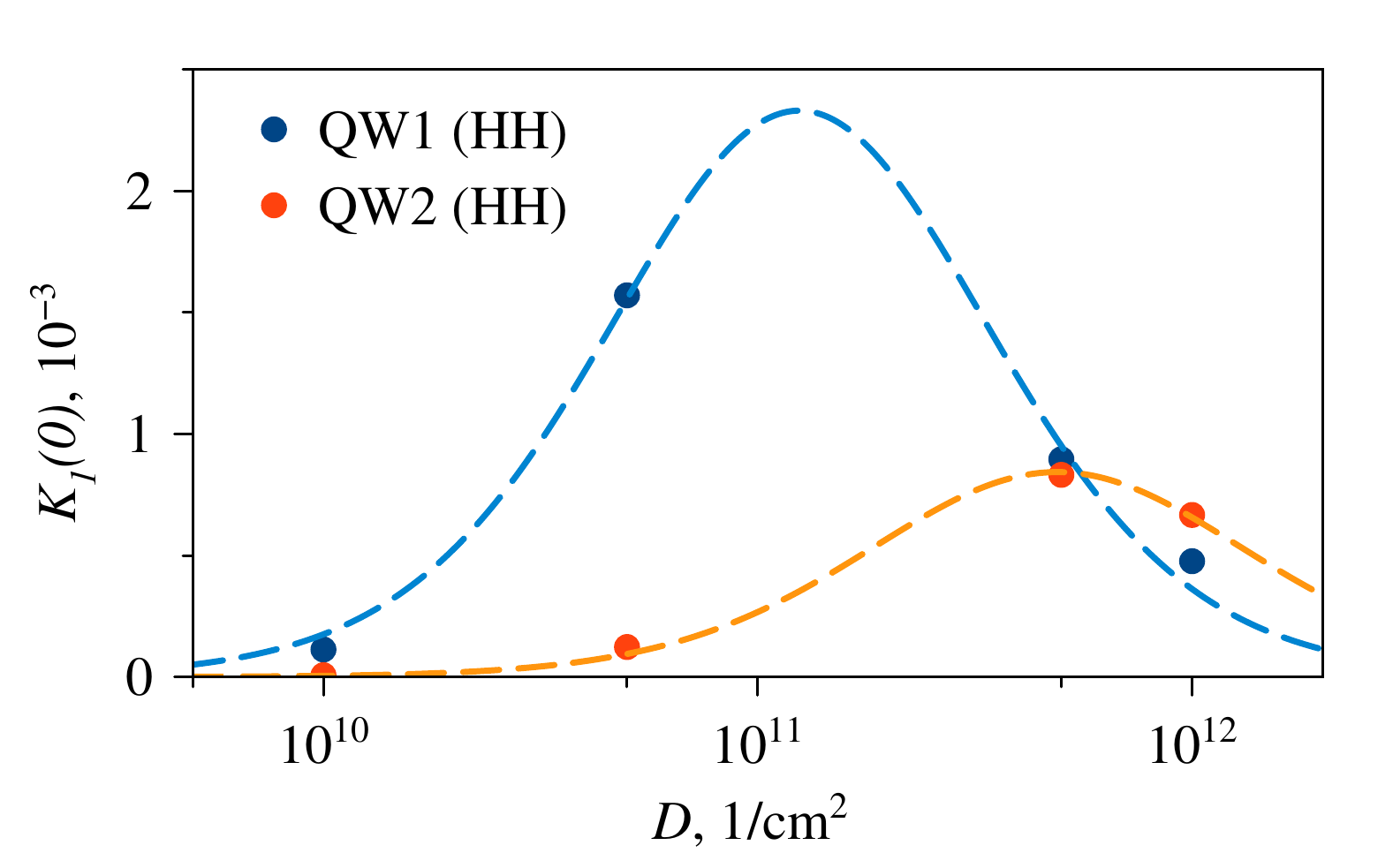}
    \caption{
    Experimental dependency of the resonant diffraction efficiency $K_1(0)$ on the ion irradiation dose $D$ for QW1~(HH) (blue dots) and QW2~(HH) (red dots) resonances. Corresponding dashed lines shows fit by~(\ref{K1inh}). Note the logarithmic dose scale.
    }
    \label{FigP602_K1D}
    \end{figure}

    Dots in Fig.~\ref{FigP602_K1D} shows dependency of the resonant diffraction efficiency $K_1(0)$ on the ion irradiation dose $D$ for QW1~(HH) and QW2~(HH) resonances. These dependencies reach a maximum at a certain dose. Previous studies have shown that ion-beam-induced inhomogeneous broadening is proportional to the irradiation dose~\cite{Kapitonov2015}. This proportionality allows us to fit the data using developed theory assuming the presence of parasitic modulation of unirradiated grating grooves (denoted above by constant $\beta$). Such fittings are shown by dashed lines.
    
    Although modeling of the ion scattering for the case of 35~keV He$^+$ ions shows approximately the same vacancy generation yield for QW1 and QW2~\cite{Kapitonov2015}, a direct comparison of the doses at which the maximum diffraction efficiency is observed does not allow us to make a conclusion about the ratio between coefficients $\beta$ for these two QWs. However, if we assume that the proportionality coefficients between the inhomogeneous broadening and irradiation dose for QW1 and QW2 are close, then a lower optimal dose in the case of QW1 indicates a greater parasitic modulation, which is consistent with the wider scattering of ions in the sample at QW1 depth.

\section{Conclusions}

    In summary, a theoretical model for excitonic resonant diffraction gratings was developed based on the step-by-step approximation of the Maxwell equation solution. The cases of spatial modulation of the frequency of the exciton resonance in the quantum well, its radiation width, or nonradiative broadening were considered. For the latter case, ways were suggested to increase the resonant diffraction coefficient by optimizing the fill factor and taking into account the proximity effect when choosing the magnitude of the spatial modulation. The theoretical model was used to explain the spectral and dose dependencies in the cases of pre- and post-MBE ion-beam-irradiated quantum wells.
    
    Sub-wavelength spatial modulation of solely resonant optical properties of quantum wells and other objects with exciton resonances is a promising method for creating various resonant diffractive optical elements (DOEs), including the resonant diffraction gratings considered in this paper. Such modulation can be used for laser beam shaping, wavelength-division multiplexing, and total-internal reflection waveguides coupling.

\section*{Funding}
Russian Science Foundation (17-72-10070).\\

\section*{Acknowledgments}
The experiment was carried out using
equipment of the SPbU Resource Centers ''Nanophotonics'' and ''Nanotechnology''.


\end{document}